\documentclass[aps,twoside,a4paper,12pt]{revtex4-1}

\usepackage{epsfig}
\usepackage{amsmath,amssymb}
\usepackage{color}

\usepackage[english]{babel}

\parskip=\medskipamount

\newcommand{\eq}[1]{(\ref{#1})}
\newcommand{\fig}[1]{Fig. \ref{#1}}

\newcommand{\be}{\begin{equation}}
\newcommand{\ee}{\end{equation}}

\newcommand\disp{\displaystyle}

\newcommand{\la}{\left<}
\newcommand{\ra}{\right>}

\def\runninghead#1#2{\pagestyle{myheadings}
\markboth{{\protect\sl{\quad #1}}\hfill} {\hfill{\protect\sl{#2\quad}}}}

\begin{document}

{\flushleft ITEP Preprint No.: ITEP--TH--34/11}
\bigskip

\runninghead{\sl A. Gorsky, S.K. Nechaev, R. Santachiara, G. Schehr}{\sl Random ballistic growth
and diffusion in symmetric spaces}

\title{Random ballistic growth and diffusion in symmetric spaces}

\author{A. Gorsky$^1$, S. Nechaev$^2$, R. Santachiara$^{2,3}$, G. Schehr$^4$}

\affiliation{$^1$ITEP, B. Cheryomushkinskaya 25, Moscow, Russia \\
$^2$LPTMS, Universit\'e Paris Sud, 91405 Orsay Cedex, France \\
$^3$J.-V. Poncelet Laboratory, Independent University of Moscow, 11 B. Vlasievsky per., 119002
Moscow, Russia \\
$^4$LPT, CNRS - Universit\'e Paris Sud, 91405 Orsay Cedex, France}

\begin{abstract}
Sequential ballistic deposition (BD) with next--nearest--neighbor (NNN) interactions in a
$N$--column box is viewed a time--ordered product of $N\times N$--matrices consisting of a single
$sl_2$--block which has a random position along the diagonal. We relate the uniform BD growth with
the diffusion in the symmetric space $H_N=SL(N,R)/SO(N)$. In particular, the distribution of the
maximal height of a growing heap is connected with the distribution of the maximal distance for the
diffusion process in $H_N$. The coordinates of $H_N$ are interpreted as the coordinates of
particles of the one--dimensional Toda chain. The group--theoretic structure of the system and
links to some random matrix models are also discussed.
\end{abstract}

\maketitle

\tableofcontents

\section{Introduction}
\label{sect:1}

During recent years much effort has been devoted to theoretical analysis of properties of surfaces
obtained by aggregation of particles. Several models describing various properties of clusters
grown by different deposition processes deserved special attention. Among them we can mention the
models of surfaces grown by Molecular Beam Epitaxy (MBE) (see, for example, Ref. \cite{MBE}),
Polynuclear Growth (PNG) \cite{M,PNG,BR,J} and by Ballistic Deposition (BD)
\cite{Mand,MRSB,KM,BMW,KTKR,MN}. All these models belong to the Kardar--Parisi--Zhang (KPZ)
universality class \cite{KPZ}. Our attention is focused on ballistic deposition. In the simplest
setting, one can assume that elementary units (``particles'') follow ballistic trajectories with
random initial positions and specified direction in space and adhere sequentially to a growing
aggregate. Despite the extremely transparent geometric formulation, the problem of stochastic
growth in the ballistic deposition process remains one of the most challenging subjects in
statistical mechanics.

The definition of a standard one--dimensional ballistic deposition model with
next--nearest--neighboring (NNN) interactions (for more details, see Refs. \cite{scaling1,
scaling2, scaling3}) is as follows. Consider a box divided in $N$ columns (of unit width each)
indexed by an integer $i$ ($i=1,2,..., N$). The free conditions are assumed for left and right
boundaries. At the initial time moment, ($t=0$), the system is empty. Then, at each tick of the
clock, $t=1,2,...,T$, we deposit an elementary cell (``particle'') of unit height and width in a
randomly chosen column, $i$. Suppose that the distribution of these nucleations on the set of
columns is uniform. Define the height, $h_i(t)$, in the column $i$ at time moment $t$. Assume now,
that the cells in the nearest--neighboring columns interact in such a way that they can only touch
each other by corners, but never by their vertical sides. This implies that after having deposited
a particle to the column $i$, the height of this column is modified according to the following
rule:
\be
h_i(t+1)=\max[h_{i-1}(t),\, h_i(t),\, h_{i+1}(t)]+1 \;.
\label{eq:1}
\ee
If at the time moment $t$ nothing is added to the column $i$, its height remains unchanged:
$h_i(t+1)= h_i(t)$. A set of deposited particles forms a heap (a pile) as it is depicted in
\fig{fig:heap} b.

The available theoretical analysis of stochastic deposition focuses mainly on the properties of the
enveloping surface $h_i(T)$. Recall the well--known scaling relation characterizing the surface of
a growing aggregate:
\be
\left[{\rm Var}\,h(T)\right]^{1/2}=\frac{1}{N^{1/2}} \left[\sum_{i=1}^{N} \la h^2_i(T) \ra - \la
h_i(T) \ra^2 \right]^{1/2} = N^{1/2}g\left(\tau/N^{3/2}\right),
\label{eq:2}
\ee
where the brackets denote an average over different realizations of the noise, the variable
$\tau=T/N$ is the averaged number of particles per one column, and the function $g(u)$ is the
rescaled variable $u=\tau/N^{3/2}$ has the following asymptotic behavior: $g(u)\sim u^{1/3}$ for
$u\ll 1$ and $g(u)\sim {\rm const}$ for $u\gg 1$. Such a behavior is typical for many
non-stationary growth problems. The exponent $1/3$ has been found already in \cite{KPZ} for the
original Kardar--Parisi--Zhang (KPZ) model and then observed in a variety of other growth models.
In \cite{johansson,spohn} it was realized that, for flat initial conditions, the distribution of a
rescaled PNG surface height, $\tau^{-1/3}(h(i, \tau)-2\tau)$, converges as $\tau\to\infty$ to the
Tracy--Widom (TW) distribution \cite{tw}, describing the edge states of random matrices belonging
to the Gaussian Orthogonal Ensemble (GOE). Note that in the droplet geometry the statistics of
growing surface is instead described by the edge states of the Gaussian Unitary Ensemble (GUE)
\cite{spohn}. Height fluctuations in the KPZ universality class were measured in experiments, both
in planar \cite{miettinen, takeuchi1, takeuchi2} and in curved geometry in the electroconvection of
nematic liquid crystals \cite{takeuchi1, takeuchi2} and a good quantitative agreement with TW
distributions was found.

The typical analytic analysis of a growth process described by the equation \eq{eq:1} consists of a
few heuristic steps which are usually beyond a precise control. First of all, one sends the lattice
spacing and time step to zero (with proper rescaling of all relevant variables), thus arriving at a
continuous nonlinear partial differential equation of the type
\be
\partial_t h(x,t) = \nu \partial^2_x h(x,t) + g \left|\partial_x h(x,t) \right| + \eta(x,t),
\label{eq:3}
\ee
where $\nu$ and $g$ are some constants and $\eta(x,t)$ is a Gaussian white noise, which replaces
the telegraph noise in the initial equation \eq{eq:1}. The next step is to replace the nonanalytic
nonlinearity $\left|\partial_x h(x,t) \right|$ by the analytic one $\left(\partial_x
h(x,t)\right)^2$. This replacement is based on numerical evidence \cite{amar} and solid
mathematical estimates \cite{GGK} which show that the universality class of KPZ equation is
insensitive to the value of $\alpha$ ($1\le \alpha \le 2$) for the nonlinearity of the type
$\left|\partial_x h(x,t) \right|^{\alpha}$.

Another point of view on ballistic deposition process has been developed recently in \cite{khanin},
where it has been demonstrated that the discrete stochastic equation describing the BD process in
\eq{eq:1} can be naturally represented in terms of a ``dynamic programming'' language associated
with the so-called Bellman equation \cite{bellman}. This ``dynamic programming'' point of view
allowed for a systematic translation of evolution of growing BD clusters in the NNN BD model into
the language of maximizers and shocks in discrete equations of the Burgers or Hamilton--Jacobi
type. This approach allowed to determine the exponents characterizing: i) the decay of the number
density of clusters, ii) the wandering of the inter--cluster interface, and iii) the tail of the
cluster mass distribution.

In the present paper we consider the ballistic deposition process from a completely different point
of view, linking it to the diffusion in some symmetric space. The goal of our work is two--fold: \\
\indent 1. On one hand, we conjecture that the uniform NNN ballistic growth in a scaling limit can
be interpreted as multiplication of random matrices of type \eq{eq:11}, which, in turn is linked to
the diffusion in the symmetric space $H_N=SL(N,R)/SO(N)$. In particular, the distribution of the
maximal height of a growing heap is connected with the distribution of the maximal distance
for the diffusion process in $H_N$; \\
\indent 2. On the other hand, we demonstrate that the coordinates of $H_N$ can be interpreted as
the coordinates of particles of the one--dimensional Toda chain, which itself can be regarded as
Inozemtsev limit of Calogero--type systems describing the transmission in random wires.

The paper is structured as follows. In the Section \ref{sect:2} we remind the concept of ``local
groups'' and demonstrate that the BD process can be described in terms of matrix multiplications;
in the Section \ref{sect:3} we describe the NNN BD process as diffusion in an appropriate symmetric
space, derive the connection with the one--dimensional Toda system and show that BD can be
described by the Inosemtsev limit of hyperbolic Calogero--type systems; in the Section
\ref{sect:4.1} we interpret some group--theoretic facts concerning Toda chain in terms of ballistic
deposition; the Conclusion summarizes our results. The two Appendices outline: a) the connection
between heights and eigenvalues, and b) the link between 1D Toda system and rational solutions of
Painlev\'e II.

\section{Ballistic deposition and products of random matrices}
\label{sect:2}

\subsection{Ballistic growth and directed polymers in random environment}
\label{sect:2.1}

Consider the set of matrices $\{g_1,...,g_N\}$, where the matrix $g_i$ has the following form:
\be
g_i=\left(\begin{array}{ccccc} 1 & & & & \\ & \ddots & & & \\
& & \boxed{\begin{array}{ccc} 1 & 0 & 0 \\ u & u & u  \\ 0 & 0 & 1 \end{array}} & & \\
& & & \ddots & \\ & & & & 1 \end{array}\right)\leftarrow \mbox{row $i$}.
\label{eq:4}
\ee
The deposition event in the column $i$ means the application of the generator $g_i$. Rising a heap
by sequential addition of particles with NNN interactions is translated into a sequential
multiplication of matrices of type \eq{eq:4}. Thus, the whole heap is represented by the
time--ordered product
\be
V(T,u)=\, :\hspace{-3pt}\prod_{t=1}^{T} g_{i_t}\hspace{-3pt}:\, .
\label{eq:5}
\ee
Each element of the matrix $V(T,u)$ is a polynomial of the variable $u$. Take a vector ${\bf
a}(t=0)=(a_1,...,a_N)$ where $a_i (i=1,...,N)$ are strictly positive values. The set of local
heights ${\bf h}(T)=(h_1(T),...,h_N(T))$ at time $T$ after deposition event can be extracted as
follows:
\be
{\bf h}(T) = \lim_{u\to\infty} \frac{\ln [V(T,u)\, {\bf a}(t=0)]}{\ln u}.
\label{eq:6}
\ee
The definition of heights in \eq{eq:4}--\eq{eq:6} is consistent with the updating rules \eq{eq:1}
for the NNN--ballistic deposition process. Namely, write the recursion for the updating dynamics of
the vector ${\bf a}(t)$:
\be
\left(\begin{array}{l} a_1(t+1) \\ \vdots \medskip \\ a_{j-1}(t+1) \\ a_j(t+1) \\ a_{j+1}(t+1)
\medskip \\ \vdots \\ a_N(t+1) \end{array}\right)=
\left(\begin{array}{ccccc} 1 & & & & \\ & \ddots & & & \\
& & \boxed{\begin{array}{ccc} 1 & 0 & 0 \\ u & u & u  \\ 0 & 0 & 1 \end{array}} & & \\
& & & \ddots & \\ & & & & 1 \end{array}\right) \left(\begin{array}{l} a_1(t) \\ \vdots \medskip \\
a_{j-1}(t) \\ a_j(t) \\ a_{j+1}(t) \medskip \\ \vdots \\ a_N(t)\end{array}\right).
\label{eq:7}
\ee
If $i_t=i$ (i.e. we drop a particle in the column $i$ ar time $t$), then
\be
a_i(t+1) = u\, a_{i-1}(t) + u\, a_i(t) + u\, a_{i+1}(t).
\label{eq:8}
\ee
Otherwise $a_i(t+1)=a_i(t)$. Substituting the ansatz $a_i(t) = e^{\beta h_i(t)}$ into \eq{eq:8}, we
get for deposition event in the column $i$:
\be
h_i(t+1) = \frac{1}{\beta}\ln\left[e^{\beta h_{i-1}(t)} + e^{\beta h_i(t)} + e^{\beta
h_{i+1}(t)}\right] + 1.
\label{eq:9}
\ee
For all other columns $j\neq i$ the height remains unchanged: $h_j(t+1)=h_j(t)$ ($j=1,...,N$). The
initial configuration by definition is ${\bf h}(T)=(0,...,0)$. Taking in \eq{eq:9} the limit
$\beta\to \infty$, we recover \eq{eq:1}. Note, that \eq{eq:9} resembles the ultra--discretization
of stochastic growth equations \cite{ud}.

\begin{figure}[ht]
\epsfig{file=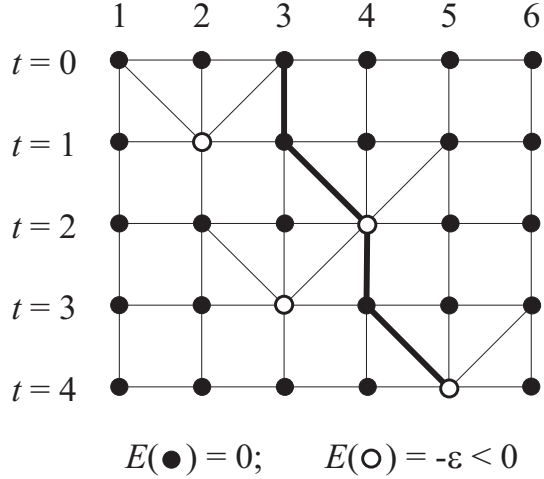,width=7cm} \caption{Directed polymer model in a disordered environment.
Open dots designate the randomly placed sites with energy $-\epsilon$.}
\label{fig_std_ballistic}
\end{figure}

The quantity $a_i(t)$ can be interpreted as a {\em finite temperature} partition function of a
directed polymer in a random environment, $\beta=\log u$ being essentially the inverse temperature.
To demonstrate this consider the random lattice depicted in the \fig{fig_std_ballistic}. The rules
of the deposition process mean that at each ``time'' $t$ a single site (an open dot) is chosen: an
energy $-\epsilon$ is assigned to this site and it is connected to three neighboring ancestors at
time $t-1$. To the other sites (black dots) an energy $0$ is assigned and they are connected
(vertically) to a single site. On this random lattice we then consider a configuration of a
directed polymer, shown by a bold line.

The energy assigned to a polymer configuration is the sum of site energies through which the
directed polymer is passing. If we denote by $e^{\beta \epsilon} = u$, then the quantity $a_i(t)$
evolving according to \eq{eq:7} gives exactly the partition function of a polymer ending in site
$i$ at time $t$. By this correspondence a precise connection between the BD a directed polymer in a
random medium is established (see, for instance, \cite{satya} and references therein).

\subsection{Ballistic deposition and random walk on local groups}
\label{sect:2.2}

We have seen that the ballistic deposition process has a natural description in terms of
multiplication of random matrices with special ``local'' structure. The established connection is a
particular example of a uniform Markov process on a ``local group'' \cite{vershik1}.

\noindent {\bf Definition 1.} {\em Local group (semi--group) $L_N$ ($L_N^+$) with $N$ generators
$g_1,...,g_N$ is a group (semi--group) with the commutation relations
\be
g_k g_m = g_m g_k,\qquad \forall |k-m|\ge 2,
\label{eq:10}
\ee
and might have additional relations $R$ inside the elementary blocks $g_k, g_{k\pm 1}$
$(k=1,...N)$. If the relations $R$ are the same for all $k$, then $L$ is called a local stationary
group \cite{vershik1}.}

Plenty of important groups, semi--groups and algebras are of type of local groups: the Coxeter
groups, Hecke algebra, braid groups, Temperley--Lieb algebra, etc. A locally free group $F$ with
$M$ generators, being in some sense the ``simplest'' nontrivial example of a local group, can be
considered as a universal object in the manifold of all local groups with a given number of
generators.

\noindent {\bf Definition 2.} {\em Locally free group (semi--group) $F_N$ ($F_N^+$) with $N$
generators $g_1,...,g_N$ is a group (semi--group), determined only by the relation \eq{eq:10}. Each
pair of neighboring generators $g_k, g_{k\pm 1}$ produces a free subgroup (sub--semigroup) of $F_N$
($F_N^+$) \cite{nech-ver}.}

\begin{figure}[ht]
\epsfig{file=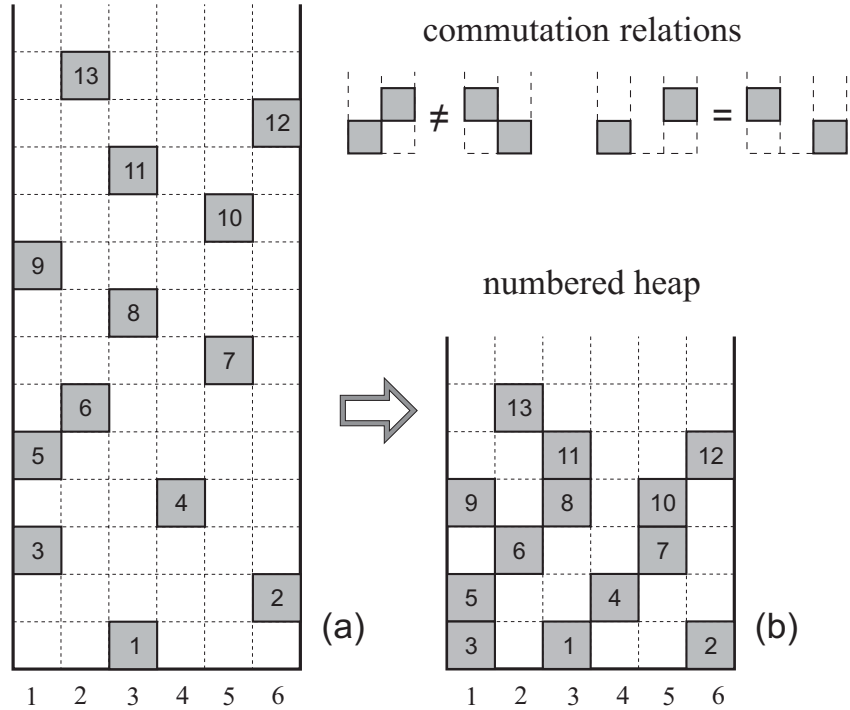, width=11cm} \caption{(a) The sequential ballistic deposition; (b) the
heap obtained by contraction procedure using commutation relations of $F_N^+$ ($N=6$).}
\label{fig:heap}
\end{figure}

Let us define the numbered heap as a heap whose cells are enumerated by sequential ``ticks of
time'', $t$. Deposition of a particle in a column $i$ means the application of a generator $g_i$.
There is a bijection between numbered heaps of $T$ sequentially deposited cells and words of length
$T$ in $F_N^+$ such that a given numbered heap is uniquely represented by a word in $F_N^+$ (and
vice versa) \cite{nech-ver}. The key point of this bijection is the special way of representing
words in the locally free semigroup (group). Namely, all words are written in the so-called
``normal order form'' implying that the generators with smaller indices are pushed as left as
possible when it is allowed by commutation relations \eq{eq:10}. For example, the word associated
with subsequent deposition of all 13 particles in \fig{fig:heap} is: $ W=g_3\, g_6\, g_1\, g_4\,
g_1\, g_2\, g_5\, g_3\, g_1\, g_5\, g_3\, g_6\, g_2$. The same word $W$ written in the normal order
form $ W=g_1\, g_1\, g_3\, g_2\, g_1\, g_4\, g_3\, g_3\, g_2\, g_6\, g_5\, g_5\, g_6$, is in
bijection with the numbered heap (\fig{fig:heap}b) obtained by ``contraction'' of the sequential
deposition (\fig{fig:heap}a). The statistical properties of locally free semigroups and groups are
investigated in details in \cite{nech-ver}. Thus, the uniform growth of a random heap on a set of
$N$ columns can be considered as a uniform Markov chain on a semigroup $F_N^+$ with $N$ generators.

It should be noted that the concept, equivalent to the locally free semi--group $F_N^+$,
appeared earlier \cite{viennot} in the investigations of combinatorial properties of substitutions
of sequences and so called ``partially commutative monoids'' \cite{cartier-foata}. Especially
productive is the geometrical interpretation of monoids in the form of a ``heap'', offered by X.G.
Viennot \cite{viennot}. It is worth mentioning that the concept of a monoid--heap appears in
connection with various aspects of directed growth, parallel computing and complexity algorithms
\cite{complexity}.

\noindent {\bf Definition 3.} {\em Take a locally free semi--group $F_N^+$ with fixed finite number
of generators $\{g_1,...,g_N\}$. Consider the uniform distribution on the set $\{g_1,..., g_N\}$.
Construct the (left--handed) Markov chain on $F_N^+$: ${\bf a}(t=0)=\{a_1,...,a_N\}$, ${\bf
a}(t)=g_{i_t}{\bf a}(t-1)$, where $i_t=\{1,...,N\}$ is uniformly distributed on the set of indices
$\{1,...,N\}$.}

\begin{figure}[ht]
\epsfig{file=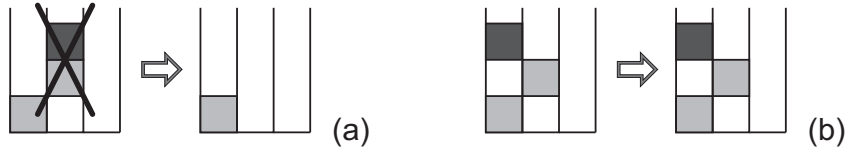, width=11cm} \caption{Deposition rules for colored heap (annihilation is
allowed).}
\label{fig:color}
\end{figure}

Analogously with locally free semi--group, we can define the Markov chain on a locally free group.
The geometric interpretation of random walk on locally free group is very straightforward. Consider
two types of particles: black and white, and identify the generators $g_i$ and $g_i^{-1}$
respectively with deposition of black and white particle in the column $i$ ($i=1,...,N$). The
elementary depositions are depicted in \fig{fig:color}, where (a) corresponds to the deposition
$g_1 g_2 g_2^{-1} \equiv g_1$ (since $g_i g_i^{-1}=1$), and the configuration shown in (b) cannot
be reduced.

\subsection{``Soft deposition''}
\label{sect:2.3}

Deposition process viewed as a random walk on a locally free semigroup (or group) can be considered
from the point of view of diffusion in symmetric spaces. Let us demonstrate this on two simple
examples: sequential deposition in two and three columns.

\begin{figure}[ht]
\epsfig{file=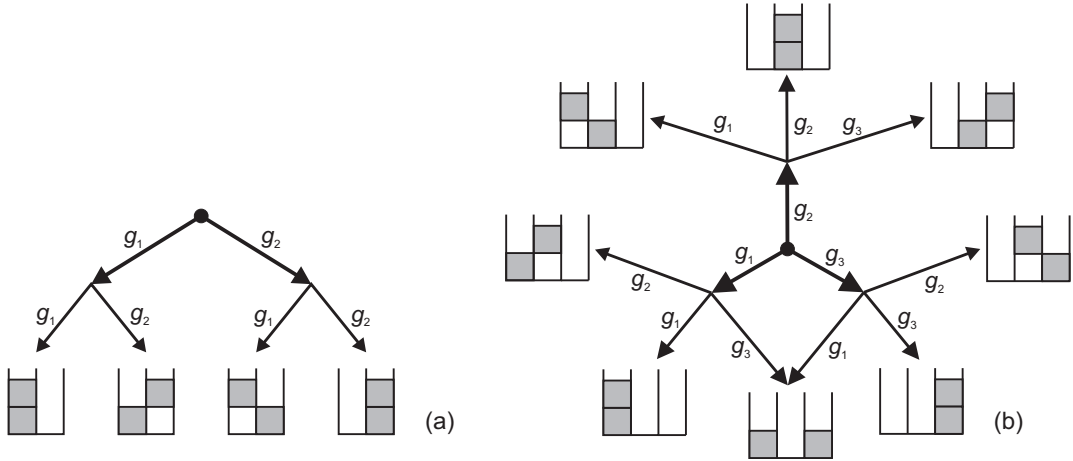, width=14cm} \caption{Ballistic deposition in two (a) and three (b)
columns labelled by states of Cayley graphs of corresponding locally free semigroups.}
\label{fig:two_three}
\end{figure}

Deposition in two columns corresponds to a Markov chain on a locally free semigroup $F_2^+$ with
two generators, $g_1, g_2$ can be represented by a Cayley tree shown in \fig{fig:two_three} a,
since for $N=2$ one has $F_2^+ \equiv \Gamma_2^+$, where $\Gamma_2^+$ is a free semigroup with two
generators (see, for example, \cite{series} for definitions). Deposition in three columns can be
treated similarly as a Markov chain on the Cayley graph associated with locally free semigroup
$F_3^+$: $\{g_1, g_2, g_3\}$ and the commutation relation $g_1 g_3 = g_3 g_1$ -- see
\fig{fig:two_three} b. Certainly, the same correspondence can be established between the ballistic
deposition with annihilation (as shown in \fig{fig:color}) and random walk on the full locally free
group $F_N$.

We shall also consider a random walk on the full group $F_N$, i.e. the uniform growth in a
two--color heap in which direct deposition of white particle on top of black one (and vice versa)
leads to their mutual annihilation.

The key problem connecting the $N$--column deposition and diffusion in symmetric space is the
following. For the multi--column deposition we need to find a metric space, $H_N$ in which the
locally free group $F_N$ with $N$ generators acts as a group of isometries, i.e. in which the
Cayley graph ${\cal C}_N$ of $F_N$ can be isometrically embedded.

It is well known that any regular Cayley tree, ${\cal C}$, as an exponentially growing structure,
cannot be isometrically embedded in an Euclidean plane. The embedding of a Cayley tree ${\cal C}$
into the metric space is called ``isometric'' if ${\cal C}$ covers that space, preserving all
angles and distances. For example, the rectangular lattice isometrically covers the Euclidean plane
$E=\{x,y\}$ with the flat metric $ds^2=dx^2+dy^2$. In the same way, the Cayley tree ${\cal C}$
isometrically covers the surface of constant negative curvature (the Lobachevsky plane) ${\cal H}$.
One of possible representations of ${\cal H}$ is the Poincar\'e model, the upper half--plane ${\rm
Im}\,z>0$ of the complex plane $z=x+iy$ endowed with the metric $ds^2=\frac{dx^2+dy^2}{y^2}$ of
constant negative curvature. Thus, the locally free group $F_2\equiv \Gamma_2$ is the group of
isometries of Lobachevsky plane. Since the Lobachevsky plane is the coset space $H_2 =
SL(2,R)/SO(2)$, we can consider the two--column ballistic deposition (with possible cancellation)
as a random walk in $H_2$.

Similarly, the locally free group $F_3$ generated by $g_1, g_2, g_3$ with the commutation relation
$g_1g_3=g_3 g_1$ acts as a group of isometries of the Lobachevsky space with the metric
$ds^2=\frac{dx^2+dy^2+dz^2}{y^2}$. Thus, the deposition in three columns has an important
difference with respect to the two--columnar growth. Namely, the steps along the coordinates $x$
and $z$ in Lobachevsky space commute which is straightforwardly translated into the commutation
relation $g_1g_3=g_3 g_1$ in the group $F_3$. Hence, the three--columnar growth corresponds to the
symmetric random walk in Lobachevsky space $H_3=SL(3,R)/SO(3)$. The consideration of the
corresponding diffusion equations in curved geometry is the subject of the next Section.

To exploit the power of algebraic methods, we suggest the following smoothed description of
$N$--columnar ballistic growth.

\noindent {\bf Definition 4.} {\em We define the ``soft'' deposition generator, $\tilde{g}_i$, as
follows:
\be
\tilde{g}_i(t)=\left(\begin{array}{ccccc} 1 & & & & \\ & \ddots & & & \\
& & \boxed{U_i^{(t)} \hspace{-1mm}} & & \\ & & & \ddots & \\ & & & & 1 \end{array}\right); \qquad
U_i^{(t)} = \left(\begin{array}{cc} a_i^{(t)} & b_i^{(t)} \medskip \\ c_i^{(t)} & d_i^{(t)}
\end{array}\right); \qquad a_i^{(t)} d_i^{(t)} - b_i^{(t)} c_i^{(t)}=1 \;,
\label{eq:11}
\ee
where the block $U_i^{(t)}$ is an $SL(2,R)$ matrix. For definiteness, suppose that the entries
$a_i^{(t)}, b_i^{(t)}, c_i^{(t)}, d_i^{(t)}$ are random variables (up to the condition $\det
U_i^{(t)}=1$) distributed in a compact support with some measure $P(u)$.}

Note that this ``smoothed'' description captures all principal properties of the growing heap which
we would like to analyze:

i) $\tilde{g}_i(t_1) \tilde{g}_j(t_2)= \tilde{g}_j(t_1) \tilde{g}_i(t_2)$ for $|i-j|\ge 2$ and
$t_2>t_1$, meaning that sequential deposition in distant (i.e. non--nearest--neighboring) columns
commutes;

ii) $\tilde{g}_i(t_1) \tilde{g}_j(t_2)\neq \tilde{g}_j(t_1) \tilde{g}_i(t_2)$ for $|i-j|=1$ and
$t_2>t_1$, meaning that sequential deposition in nearest--neighboring columns does not commute.

The group $\tilde{G}: \{\tilde{g}_1, ...,\tilde{g}_N\}$ falls under the Definition 1 of local
groups. The advantage of the redefinition of deposition generators $g_i\to \tilde{g}_i$ [compare
\eq{eq:4} and \eq{eq:11}] will appear below. Equipped with these definitions, we formulate the
following conjecture:

\noindent{\bf Conjecture.} {\em The Cayley graphs associated to the locally free group generated
by $g_i$'s on one hand, and to the local group generated by $\tilde{g}_i$'s on the other hand, can
both be embedded isometrically in the same metric space $H_N$.}

The metric space $H_N$ is the symmetric space $H_N = SL(N,R)/SO(N)$. One can then use the Lie group
structure of this symmetric space to define the Beltrami--Laplace operator which coincides with the
Casimir operator $C_2$. This would allow us to describe the symmetric Markov chain on the locally
free group $F_N$ with $N$ generators by the uniform diffusion process in the symmetric space $H_N$,
where the group $F_N$ acts as a group of isometries.

The connection between the columnar heights, $h_j(t)$ (defined in \eq{eq:9}), and current
coordinates $\mu_j(t)$ describing the diffusion process in the symmetric space $H_N$, is as
follows. Denote by $\tilde{V}_t$ the time--ordered product of $t$ generators $\tilde{g}$ in
\eq{eq:11}. Let $\{\lambda_1(t),..., \lambda_N(t)\}$ be the set of eigenvalues of the matrix
$\tilde{V}_t$; define $\mu_j(t) = \ln \lambda_j(t)$ ($j=1,...,N$). Then, one has for $t\gg 1$:
\be
h_{\rm max}(t) = \gamma_{N}\; \mu_{\rm max}(t)
\label{eq:hyp_dist1}
\ee
where $h_{\rm max}(t)=\max \{h_1(t),...,h_N(t)\}$, $\mu_{\rm max}(t) = \max \{\mu_1(t),
...,\mu_N(t)\}$, and $\gamma_{N}$ is some constant which depends only on the matrix size, $N$. Our
numerical computations show that $\gamma_N$ converges rapidly to its limiting value $\gamma\approx
0.8$ as $N$ increases -- see Appendix \ref{app:1}. We also describe in Appendix \ref{app:1} the
relation (for $N=2$) between the number of steps, $n$, from the origin along a Cayley tree embedded
into $H_2=SL(2,R)/SO(2)$, and a hyperbolic distance $d$ in $H_2$.

\section{Diffusion in symmetric spaces}
\label{sect:3}

\subsection{Sequential deposition in two columns and a diffusion in $H=SL(2,R)/SO(2)$}
\label{sect:3.1}

For $N=2$ the sequential deposition process is viewed as a multiplication of $SL(2,R)$--random
matrices. The equation for the stochastic evolution of the vector ${\bf r}(t) = (x(t), y(t))$ is
\be
\left(\begin{array}{c} x(t+1) \medskip \\ y(t+1) \end{array} \right) = \left(\begin{array}{cc}
a^{(t)} & b^{(t)} \medskip \\ c^{(t)} & d^{(t)} \end{array}\right) \left(\begin{array}{c} x(t)
\medskip \\ y(t) \end{array} \right).
\label{eq:12}
\ee
which can be conveniently rewritten, introducing the Ricatti variable $\rho(t)=x(t)/y(t)$, as:
\be
\rho(t+1) = \frac{a^{(t)}+b^{(t)} \rho(t)}{c^{(t)}+d^{(t)} \rho(t)}.
\label{eq:13}
\ee
The stochastic equation \eq{eq:12} was the subject of the pioneering work \cite{gerts} where it has
been explicitly shown that the limiting distribution function $W_T(\rho)$ for $T\gg 1$ and
$a_i^{(t)}, b_i^{(t)}, c_i^{(t)}, d_i^{(t)}$ randomly distributed on a compact support with one and
the same distribution for all $t=1,...,T$, satisfies the diffusion equation in the Lobachevsky
plane (the two--dimensional Riemann surface of constant negative curvature). Recall that the
Lobachevsky plane coincides with the symmetric space $H_2=SL(2,R)/SO(2)$. So, we could immediately
write the heat equation in $H_2$ for the distribution function $P({\bf x},t)$, where $t$ is the
``time'' and ${\bf x}$ are the coordinates of a Riemann manifold:
\be
\partial_t P({\bf x},t) = D \Delta_{\rm BL} P({\bf x},t); \quad P({\bf x},t=0)=\delta({\bf
x})
\label{eq:lob1}
\ee
where
\be
\Delta_{\rm BL}=\frac{1}{\sqrt{g}}\frac{\partial}{\partial x_i} \left(\sqrt{g}
\left(g^{-1}\right)_{ik}\frac{\partial}{\partial x_k}\right)
\label{eq:lob2}
\ee
is the Beltrami--Laplace operator and $g_{ik}$ is the metric tensor of the manifold; $g=\det
g_{ik}$. The function $P({\bf x},t)$ satisfies the normalization condition $\int \sqrt{g}\, W({\bf
x},t) d{\bf x} = 1$.

It seems to be instructive to outline the derivation of \eq{eq:lob1} using the standard technique
of extracting the Lie algebra generators from the appropriate matrix decomposition and writing the
corresponding Casimir operator. Let us note that the proof of convergence and existence of limiting
distribution associated with the stochastic equation \eq{eq:13} was a subject of the work
\cite{karp}.

The $sl_2$ Lie algebra is defined by the relations \cite{gauss}
\be
[X_1,X_2]=X_3; \quad [X_1,X_3]=-2X_1; \quad [X_2,X_3]=2X_2,
\label{eq:15}
\ee
where the generators $X_1,X_2,X_3$ have the $2\times 2$ matrix representation
\be
X_1=\left(\begin{array}{cc} 0 & 1 \\ 0 & 0 \end{array} \right); \quad X_2=\left(\begin{array}{cc} 0
& 0 \\ 1 & 0 \end{array} \right); \quad X_3=\left(\begin{array}{cc} 1 & 0 \\ 0 & -1 \end{array}
\right).
\label{eq:16}
\ee

As we shall see later, the geometry of the system presumes the use of Gauss matrix decomposition
\cite{gauss}. So, we represent the $SL(2,R)$--matrix $U$ as:
\be
U\equiv \left(\begin{array}{cc} a & b \\ c & d \end{array} \right)= \left(\begin{array}{cc} 1 & 0
\\ y & 1 \end{array} \right) \left(\begin{array}{cc} e^{\pm\mu} & 0 \\ 0 & e^{\mp\mu}
\end{array} \right) \left(\begin{array}{cc} 1 & x \\ 0 & 1 \end{array} \right).
\label{eq:17}
\ee
In the differential form (in the right representation) the $sl_2$--generators $X_1,X_2,X_3$ read:
\be
X_1=\partial_x; \quad X_2=e^{\mp 2\mu}\partial_y \pm x\partial_{\mu}-x^2\partial_x; \quad X_3=-2x
\partial_x \pm \partial_{\mu}.
\label{eq:18}
\ee
The Casimir $C_2$, which coincides with the Beltrami--Laplace operator $\Delta_{\rm BL}$, can be
expressed as follows
\be
C_2=\Delta_{\rm BL}=X_1 X_2 + X_2 X_1 + \frac{1}{2}
X_3^2=\frac{1}{2}\partial_{\mu}^2\pm\partial_{\mu}+2e^{\mp2\mu}\partial_x \partial_y.
\label{eq:19}
\ee
Thus, the diffusion equation for the distribution function $W(\mu,x,y,t)$ has the following form
\be
\partial_t W(\mu,x,y,t)= D\left(\frac{1}{2}\partial_{\mu}^2 \pm \partial_{\mu}+
2e^{\mp 2\mu}\partial_x \partial_y\right)W(\mu,x,y,t),
\label{eq:20}
\ee
where $D$ is the diffusion coefficient. Let us make a comment concerning two options to choose the
sign in equations \eq{eq:17}--\eq{eq:20}. It is obvious that in \eq{eq:17} the sign can be chosen
arbitrary, however in \eq{eq:18}--\eq{eq:20} the sign should be consistent with correct physical
behavior of the system at large times. The simplest way to chose the correct sign is to compare the
asymptotic behavior of $W$ with the limiting distribution of large number of noncommuting matrices.
The classical paper by H. F\"urstenberg \cite{fuerst} states, that under some natural conditions
typical products of $T$ noncommuting matrices [$SL(2,R)$ belongs to the class under consideration]
increase exponentially with $T$. This selects $C_2=\frac{1}{2}\partial_{\mu}^2-\partial_{\mu}+
2e^{2\mu}\partial_x \partial_y$. Seeking now the solution of \eq{eq:20} in the form $W(\mu,x,y,t) =
e^{at+b\mu}Z(\mu,x,y,t)$ and choosing $a=-\frac{D}{2},\; b=1$ we arrive at the Liouville--type
equation for the function $U(\mu,x,y,t)$:
\be
\partial_t Z(\mu,x,y,t) = \frac{D}{2}\left(\partial_{\mu}^2+4 e^{2\mu}\partial_x \partial_y\right)
Z(\mu,x,y,t).
\label{eq:21}
\ee
In what follows we are interested in the $\mu$--dependence (``radial part'') of the function $U$
only. In this case the ``angular'' dependence on $x$ and $y$ can be eliminated by the reduction
conditions \cite{marshakov}, which imply the replacement $\partial_x\partial_y \to \varkappa^2 =
{\rm const}$. Thus we arrive at the equation
\be
\partial_t Z(\mu,t) = \frac{D}{2}\left(\partial_{\mu}^2+4 \varkappa^2 e^{2\mu} \right)
Z(\mu,t),
\label{eq:21a}
\ee
subject to the initial condition $Z(\mu,0)=\delta(\mu)$. The solution of \eq{eq:21a} is well known
and can be written explicitly, for example, as in \cite{comtet}
\be
\begin{array}{l}
\disp Z(\mu,t)=\frac{1}{2\pi}\int_{-\infty}^{\infty}dk \psi_k(\mu)\psi_k^*(0) e^{-k^2 t},
\medskip \\ \disp \psi_k(\mu) = 2\sqrt{\frac{2k}{D\pi}\sinh\frac{2k\pi}{\sqrt{D}}}
K_{2ik/\sqrt{D}}\left(4\sqrt{\frac{\varkappa^2}{D}}e^{\mu}\right) \;,
\end{array}
\label{eq:liouv}
\ee
where $K_a(z)$ is the Macdonald function. For generic forms of solution of \eq{eq:21a} see
\cite{terras}.

\subsection{Sequential deposition in three columns and diffusion in $H=SL(3,R)/SO(3)$}
\label{sect:3.2}

As we have seen in Introduction, for $N=3$ the sequential deposition process is not reduced to a
random walk in $H_2$ and the situation is more involved. If one considers two sequential deposition
events, we can encounter with equal probability all configurations schematically depicted below.
Namely, besides the former $SL(2,R)$ matrix multiplication
\be
\left(\begin{array}{cc} \boxed{\begin{array}{c} \\[-4mm] sl_2 \\[1mm]
\end{array}} & \\ & 1 \end{array}\right) \left(\begin{array}{cc} \boxed{\begin{array}{c} \\
[-4mm] sl_2 \\[1mm] \end{array}} & \\ & 1 \end{array}\right); \qquad
\left(\begin{array}{cc} 1 & \\ & \boxed{\begin{array}{c} \\[-4mm] sl_2 \\[1mm]
\end{array}} \end{array}\right) \left(\begin{array}{cc} 1 & \\ &
\boxed{\begin{array}{c} \\[-4mm] sl_2 \\[1mm]
\end{array}} \end{array}\right),
\label{eq:22}
\ee
we have also the matrix multiplication of two ``shifted'' copies $SL(2,R)_{+}\otimes SL(2,R)_{-}$,
where the notation $+(-)$ designates upper (lower) $sl_2$--block
\be
\left(\begin{array}{cc} \boxed{\begin{array}{c} \\[-4mm] sl_2 \\[1mm]
\end{array}} & \\ & 1 \end{array}\right) \left(\begin{array}{cc} 1 & \\ &
\boxed{\begin{array}{c} \\[-4mm] sl_2 \\[1mm]
\end{array}} \end{array}\right); \qquad \left(\begin{array}{cc} 1 & \\ &
\boxed{\begin{array}{c} \\[-4mm] sl_2 \\[1mm]
\end{array}} \end{array} \right) \left(\begin{array}{cc} \boxed{\begin{array}{c} \\
[-4mm] sl_2 \\[1mm] \end{array}} & \\ & 1 \end{array}\right).
\label{eq:23}
\ee

It can be easily demonstrated that the multiplications of type \eq{eq:22}--\eq{eq:23} do generate
the whole group $SL(3,R)$. We use the Lie algebra representation of two ``shifted'' algebras $sl_2$
and check that {\em all} $sl_3$ generators can be obtained only from the multiplications of type
\eq{eq:22}--\eq{eq:23}. Using $sl_2$--representation \eq{eq:16} we construct the set of generators
of two ``shifted'' copies of $sl_2$:
\be
\begin{array}{lll}
X_1=\left(\begin{array}{ccc} 0 & 1 & 0 \\ 0 & 0 & 0 \\ 0 & 0 & 0 \end{array} \right); &
X_2=\left(\begin{array}{ccc} 0 & 0 & 0 \\ 1 & 0 & 0 \\ 0 & 0 & 0 \end{array} \right); &
X_3=\left(\begin{array}{ccc} 1 & 0 & 0 \\ 0 & -1 & 0 \\ 0 & 0 & 0 \end{array} \right); \medskip \\
Y_1=\left(\begin{array}{ccc} 0 & 0 & 0 \\ 0 & 0 & 1 \\ 0 & 0 & 0 \end{array} \right); &
Y_2=\left(\begin{array}{ccc} 0 & 0 & 0 \\ 0 & 0 & 0 \\ 0 & 1 & 0 \end{array} \right); &
Y_3=\left(\begin{array}{ccc} 0 & 0 & 0 \\ 0 & 1 & 0 \\ 0 & 0 & -1 \end{array} \right)
\end{array}.
\label{eq:24}
\ee
These generators satisfy the commutation relations of $sl_3$ algebra:
\be
\begin{array}{c}
\left[ X_1, X_2 \right]=X_3; \quad \left[ X_1, X_3 \right]=-2X_1; \quad \left[ X_2, X_3 \right]=2X_2;
\medskip \\ \left[ Y_1, Y_2 \right]=Y_3; \quad \left[ Y_1, Y_3 \right]=-2Y_1; \quad \left[ Y_2, Y_3
\right]=2Y_2; \medskip \\ \left[ X_1, Y_3 \right]=X_1; \quad \left[ X_2, Y_3 \right]=-X_2; \quad
\left[ Y_1, X_3 \right]=Y_1; \quad \left[ Y_2, X_3 \right]=-Y_2
\end{array}.
\label{eq:25}
\ee
However the list \eq{eq:24} does not enumerates all generators of $sl_3$: two generators of $sl_3$,
$X_4$ and $Y_4$, where
\be
X_4=\left(\begin{array}{ccc} 0 & 0 & 1 \\ 0 & 0 & 0 \\ 0 & 0 & 0 \end{array} \right); \quad
Y_4=\left(\begin{array}{ccc} 0 & 0 & 0 \\ 0 & 0 & 0 \\ 1 & 0 & 0 \end{array} \right).
\label{eq:25a}
\ee
are still missing. They can be straightforwardly recovered as
\be
X_4 =\left[ X_1, Y_1 \right]\equiv X_1 Y_1 ; \quad Y_4 =\left[ Y_2, X_2 \right]\equiv Y_2 X_2 \;.
\label{eq:26}
\ee
Thus, by using the set of two triplets $\left((X_1,X_2,X_3), (Y_1,Y_2,Y_3)\right)$ (and their
commutation relations) we generate the whole group $SL(3,R)$. In other words, the sequential matrix
multiplication of type \eq{eq:22}--\eq{eq:23} generates the group $SL(3,R)$. Qualitatively, this
can be seen by writing the square of the metric element of two copies of Lobachevsky planes,
$ds_{1,2}^2$, in the upper half--plane representation in subspaces $(x,y)$ and $(y,z)$ of the
three--dimensional half--space $(x,y>0,z)$ (Siegel space):
\be
\left\{\begin{array}{ll} ds_1^2=\frac{dx^2 + dy^2}{y^2} & \mbox{for $H_2^{+}=SL(2,R)_+/SO(2)_+$}
\medskip \\ ds_2^2=\frac{dy^2 + dz^2}{y^2} & \mbox{for $H_2^{-}=SL(2,R)_{-}/SO(2)_{-}$}
\end{array}\right.
\label{eq:27}
\ee
The square of the metric element, $ds^2=ds_{+}^2+ds_{-}^2$ in the Siegel space $(x,y>0,z)$ thus
reads
\be
ds^2=\frac{dx^2+2dy^2+dz^2}{y^2}.
\label{eq:28}
\ee
By rescaling $y\to y/\sqrt{2}$ we arrive at the standard Riemann metric of the Lobachevsky space
$H_3=SL(3,R)/SO(3)$,
\be
d\tilde{s}^2=\frac{1}{2}ds^2=\frac{dx^2+dy^2+dz^2}{y^2}.
\label{eq:29}
\ee
Equation \eq{eq:29} allows us to see that the subspaces $(x,y)$ and $(y,z)$ are two copies of
Lobachevsky planes: $H_2^{+}$ and $H_2^{-}$.

One expects that after a sufficient number of depositions, the Haar measure is smeared over the
whole group $SL(3,R)$ such that we arrive at the random walk in the symmetric space
$H_3=SL(3,R)/SO(3)$. The diffusion equation associated with \eq{eq:22}--\eq{eq:23} is governed by
the Beltrami--Laplace operator $\Delta_{\rm BL}$. The most straightforward way of constructing it
consists in writing the Casimir $C_2$ for infinitesimal transformations on $SL(3,R)/SO(3)$ (see,
for example, \cite{marshakov}). Again using the Gauss decomposition, we can decompose the generic
matrix $U\in SL(3,R)$ as follows:
\be
U\equiv \left(\begin{array}{ccc} a_{11} & a_{12} & a_{13} \medskip \\ a_{21} & a_{22} & a_{23}
\medskip \\ a_{31} & a_{32} & a_{33} \end{array} \right) = \left(\begin{array}{ccc} 1 & 0 & 0
\medskip \\ y_1 & 1 & 0 \medskip \\ y_2 & y_3 & 1 \end{array} \right) \left(\begin{array}{ccc}
e^{\mu_1} & 0 & 0 \medskip \\ 0 & e^{\mu_2-\mu_1} & 0 \medskip \\ 0 & 0 & e^{-\mu_2} \end{array}
\right) \left(\begin{array}{ccc} 1 & x_1 & x_2 \medskip \\ 0 & 1 & x_3 \medskip \\ 0 & 0 & 1
\end{array} \right).
\label{eq:30}
\ee
The infinitesimal generators of the group $SL(3,R)$ in the right representation form are as follows
\be
\begin{array}{lll}
X_1 & = & \partial_{x_1} \medskip \\
X_2 & = & e^{\mu_2-2\mu_1}\left(\partial_{y_1}+y_3\partial_{y_2}
\right)+x_1\partial_{\mu_1}-x_1^2\partial_{x_1}-x_1
x_2\partial_{x_2}-(x_2-x_1x_3)\partial_{x_3} \medskip \\
X_3 & = & -2x_1\partial_{x_1}+x_3\partial_{x_3}-x_2\partial_{x_2}+\partial_{\mu_1} \medskip \\
X_4 & = & \partial_{x_2} \medskip \\
Y_1 & = & \partial_{x_3}+x_1\partial_{x_2} \medskip \\
Y_2 & = & e^{\mu_1-2\mu_2}\partial_{y_3}+x_3\partial_{\mu_2}+x_2\partial_{x_1}-x_3^2\partial_{x_3}
\medskip \\ Y_3 & = & x_1\partial_{x_1}-2x_3\partial_{x_3}-x_2\partial_{x_2}+\partial_{\mu_2}
\medskip \\ Y_4 & = & e^{\mu_2-2\mu_1}x_3\partial_{y_1}-e^{\mu_1-2\mu_2}x_1\partial_{y_3}+
\left(e^{-\mu_1-\mu_2}+y_3x_3 e^{\mu_2-2\mu_1}\right)\partial_{y_2}+x_2\partial_{\mu_1} \medskip \\
& & -(x_1x_3-x_2)\partial_{\mu_2}-
x_1x_2\partial_{x_1}+x_3(x_1x_3-x_2)\partial_{x_3}-x_2^2\partial_{x_2}
\end{array}.
\label{eq:31}
\ee
The Casimir $C_2$ on $SL(3,R)$ has the following form:
\be
C_2=X_1 X_2 + X_2 X_1 + Y_1 Y_2 + Y_2 Y_1 + X_4 Y_4 + Y_4 X_4 + \frac{2}{3} \left(X_3^2 + Y_3^2 +
X_3 Y_3 \right).
\label{eq:32}
\ee
Writing $C_2$ explicitly in terms of the generators given in Eq. \eq{eq:31}, we get:
\be
\begin{array}{l}
\frac{1}{2}C_2 = \frac{1}{2}\Delta_{\rm BL} = \frac{1}{3}\left(\partial_{\mu_1}^2
+\partial_{\mu_2}^2+ \partial_{\mu_1}\partial_{\mu_2}\right)+\partial_{\mu_1}+\partial_{\mu_2}
\medskip \\ \hspace{0.5cm} + e^{\mu_1-2\mu_2}\partial_{x_3}\partial_{y_3} + e^{-\mu_1-\mu_2}
\partial_{x_2} \partial_{y_2} + e^{-2\mu_1+\mu_2}\left(\partial_{x_1}\partial_{y_1}+
x_3\partial_{x_2}\partial_{y_1}+y_3\partial_{x_1}\partial_{y_2}+x_3 y_3
\partial_{x_2}\partial_{y_2}\right)
\end{array}.
\label{eq:33}
\ee
Let us introduce the ``radial'' function $Z(\mu_1,\mu_2,t)$ such that $W(\mu_1,\mu_2,t) =
e^{-Dt-(\mu_1+\mu_2)}$ $\times Z(\mu_1,\mu_2,t)$ and apply to \eq{eq:33} the reduction scheme
\cite{marshakov} as in \eq{eq:21a} fixing $\partial_{x_2}= \partial_{y_2}=0$. Denoting
$\partial_{x_1}\partial_{y_1}=\varkappa_1^2$ and $\partial_{x_3}\partial_{y_3}=\varkappa_2^2$,
where the constants $\varkappa_1$ and $\varkappa_2$ do not depend on the radial coordinates $\mu_1$
and $\mu_2$, we get the diffusion equation for the radial part $Z$:
\be
\partial_t Z(\mu_1,\mu_2,t) = \frac{D}{3}\left[\partial_{\mu_1}^2+\partial_{\mu_2}^2+
\partial_{\mu_1}\partial_{\mu_2}+ 3\varkappa_1^2 e^{\mu_1-2\mu_2} +
3\varkappa_2^2 e^{-2\mu_1+\mu_2}\right] Z(\mu_1,\mu_2,t).
\label{eq:34}
\ee
The symmetry of the problem allows us to set in the equation \eq{eq:34} $\varkappa_1=\varkappa_2
\equiv\varkappa$. The corresponding invariant Riemann metric $ds^2$ is
\be
\begin{array}{lll}
\frac{1}{2}ds^2 & = &  \frac{1}{2}{\rm Tr}\left[ g^{-1} dg\, g^{-1} dg \right] \medskip \\ & = &
\partial_{\mu_1}^2 - \partial_{\mu_1} \partial_{\mu_2} + \partial_{\mu_2}^2 + e^{2\mu_1-\mu_2}
\partial_{x_1} \partial_{y_1} + e^{-\mu_1 + 2 \mu_2} \partial_{x_3}
\partial_{y_3}  + x_3 y_3 e^{\mu_1 + \mu_2} \partial_{x_1}\partial_{y_1}
\end{array}.
\label{eq:35}
\ee

As we shall see in the next Section, the diffusion equation \eq{eq:34} coincides with the diffusion
equation constructed with the one--dimensional Toda Hamiltonian for two particles in the Chevalley
basis \cite{chevalley} with eliminated motion of the center of mass of the system. Note that the
fact that we obtain the Casimir $C_2$ in \eq{eq:33} in the center--of--mass frame is the
consequence of working with $SL(2,R)$ matrices with determinant $1$.

\subsection{Sequential deposition in $N$ columns and one--dimensional Toda chain}
\label{sect:3.3}

The generic case $N\ge 3$ can be constructed by considering the algebra of $N$ shifted copies of
local Lie algebras $sl_2$. The Markov chain on the uniform distribution of $N$ elementary
$sl_2$--blocks results in growth of the random band--like matrix schematically depicted as follows:
\be
\underbrace{\left(\begin{array}{cccc} \boxed{\begin{array}{c} \\[-4mm] sl_2 \\[1mm]
\end{array}} & & & \\ & 1 & & \\ & & 1 & \\ & & & \ddots \end{array}\right)
\left(\begin{array}{cccc} 1 & & & \\ & \boxed{\begin{array}{c} \\[-4mm] sl_2 \\[1mm]
\end{array}} & & \\ & & 1 & \\ & & & \ddots \end{array}\right)
\left(\begin{array}{cccc} 1 & & & \\ & 1 & & \\ & & \boxed{\begin{array}{c} \\[-4mm] sl_2 \\[1mm]
\end{array}} & \\ & & & \ddots \end{array}\right)}_{t ~\rm matrices } \to \tilde{V}_N(t) \equiv
\left(\begin{array}{ccccc} \boxed{\begin{array}{c} \\[-3.8mm] \hspace{4.2mm} \\
\end{array}} & & & & \\[-3.8mm] & \hspace{-6.2mm} \boxed{\begin{array}{c} \\[-3.8mm] \hspace{4.2mm} \\
\end{array}} & & & \\[-3.8mm] & & \hspace{-6.4mm} \boxed{\begin{array}{c} \\[-3.8mm] \hspace{4.2mm} \\
\end{array}} & & \\ & & & & \ddots \end{array}\right).
\label{eq:36}
\ee

In order to capture the physics of the ballistic growth process it is instructive to begin with
non-rigorous, essentially intuitive, construction. Suppose for a moment that $N$ is odd and the
deposition events are such that the positions of $sl_2$--blocks either coincide, or do not overlap
at all. One particular example is shown below:
\be
\left(\begin{array}{cccc} \boxed{\begin{array}{c} \\[-4mm] sl_2 \\[1mm]
\end{array}} & & & \\ & 1 & & \\ & & 1 & \\ & & & \ddots \end{array}\right)
\left(\begin{array}{cccc} 1 & & & \\ & 1 & & \\ & & \boxed{\begin{array}{c} \\[-4mm] sl_2 \\[1mm]
\end{array}} & \\ & & & \ddots \end{array}\right)
\left(\begin{array}{cccc} 1 & & & \\ & 1 & & \\ & & \boxed{\begin{array}{c} \\[-4mm] sl_2 \\[1mm]
\end{array}} & \\ & & & \ddots \end{array}\right)...
\label{eq:36a}
\ee
In this case we can easily associate the growth process with the radial diffusion on $N/2$ {\em
noninteracting} copies of Lobachevsky plane $H_2=SL(2,R)/SO(2)$. By \eq{eq:21a} we can easily
construct the radial part of the Hamiltonian of $N$ noninteracting particles
\be
H^{(0)}=\sum_{j=1}^{N} \left(p_j^2 + \kappa e^{\mu_j}\right),
\label{eq:37}
\ee
where $p_j = \frac{\partial}{\partial \mu_j}$ is the momentum of the particle $j$ (recall that in
the deposition process $j$ is the column number), while $\mu_j$ is associated with the height in
the column $j$.

The simplest way to incorporate the interactions between nearest--neighboring columns which flatten
the growing surface, making it less rough, is to replace the Hamiltonian $H^{(0)}$ of
noninteracting particles by the following one
\be
H=\sum_{j=1}^N \left(p_j^2 + \kappa e^{\mu_j-\mu_{j+1}}\right); \quad \mu_{N+1}\equiv 0.
\label{eq:38}
\ee
We obtain this Hamiltonian by direct computations for $N=2,3$ and bring also in Section
\ref{sect:4.1} group theoretic arguments in favor of \eq{eq:38} for arbitrary $N$.

First of all, let us demonstrate that the diffusion equation \eq{eq:34} obtained by Gauss
decomposition of the group $SL(3,R)$ do coincide with the two--particle Toda Hamiltonian
$$
H(2)=p_1^2 + p_2^2 + \kappa e^{\mu_1-\mu_2},
$$
under the special choice of the coordinates. Actually, in the frame of the center--of--mass for the
$N$--body system, the variables $\mu_j$ ($j=1,...,N$) are transformed as follows \cite{elegant}
\be
\begin{array}{l}
\disp \mu_1 \to \mu_1+\frac{\mu_{N+1}}{N+1}; \quad  \mu_j \to
-\mu_{j-1}+\mu_j+\frac{\mu_{N+1}}{N+1} \;, 2 \leq j \leq N; \quad \mu_{N+1}
\to -\mu_N +\frac{\mu_{N+1}}{N+1} \medskip \\
\disp p_j \to \frac{1}{N+1}\left[\sum_{i=j}^N(N+1-i)p_{i}-\sum_{i=1}^{j-1}i
p_i \right]+p_{N+1} \;, 1 \leq j \leq N \;, \\
\disp p_{N+1} \to -\frac{1}{N+1}\sum_{i=1}^{N}i p_i+p_{N+1} \;.
\end{array}.
\label{eq:39}
\ee

The explicit form of the $N$--body Toda Hamiltonian in the coordinates \eq{eq:39} is given in
\cite{elegant}. For our purposes it is sufficient to check the $N=2$--case only. The corresponding
result is:
\be
H(2) \to \frac{1}{3}\left(p_1^2+ p_2^2 + p_1 p_2\right)+ \kappa \left(e^{\mu_1-2\mu_2} +
e^{-2\mu_1+\mu_2}\right),
\label{eq:40}
\ee
which after the substitution $p_j = \frac{\partial}{\partial \mu_j}$ gives (for $\varkappa_1^2=
\varkappa_2^2=\kappa$) the diffusion equation \eq{eq:34}.

\subsection{Toda chain and scaling limit of hyperbolic Calogero--type systems}
\label{sect:3.4}

The Markov chain corresponding to the uniform sequential deposition of elementary $sl_2$--blocks
results in a growth of a random band--like matrix $\tilde{V}_N(t)$ [as shown schematically in
\eq{eq:36}] with a time--dependent bandwidth, $b(t)$, and Gaussian distribution of matrix elements
inside the band. Let us repeat that sequential growth means that at each tick of the clock,
$t=1,2,...,T$, we multiply the current $N\times N$ matrix $U_i^{(t)}$ by an elementary
$sl_2$--block in a randomly chosen place $i$ along the diagonal [see Eq. \eq{eq:11}] with uniform
distribution. The above analysis of ``microscopic'' transformation of a matrix $\tilde{V}_N$ (from
$\tilde{V}_N(t)$ to $\tilde{V}_N(t+1)$) results in the diffusion equation for the distribution
function $W(\mu_1,..., \mu_N,t)$
\be
\frac{\partial}{\partial t}W(\mu_1,..., \mu_N,t) = D \sum_{j=1}^N \left(\frac{\partial^2}{\partial
{\mu_j}^2} + \kappa e^{\mu_j-\mu_{j+1}} \right) W(\mu_1,...,\mu_N,t); \qquad (\mu_{N+1}=0),
\label{eq:c1}
\ee
where $\mu_j=\ln \lambda_j$ ($j=1,...,N$) and $\lambda_j$ are the eigenvalues of a matrix
$\tilde{V}_N(t)$.

Suppose now that we record changes in matrix $\tilde{V}_N$ within some ``mesoscopic'' time
interval, $\Delta t$, thus, producing a sequence $\{\tilde{V}_N(t=0), \tilde{V}_N(\Delta t),
\tilde{V}_N(2\Delta t),..., \tilde{V}_N(k\Delta t)\}$ ($k\gg 1$). Choosing $\Delta t\sim N$, we
ensure the uniform growth of a bandwidth $b$ of a matrix $\tilde{V}_N$ with time. Until the width
$b$ of the band is of order of unity, one cannot say anything concerning the distribution of
eigenvalues of $\tilde{V}_N(t)$. Meanwhile, it is known (see \cite{fyodorov}) that for $b\gg
\sqrt{N}$ one can apply the results of the Dorokhov--Mello--Pereyra--Kumar (DMPK) theory
\cite{dmpk} to the band matrices. Recall that DMPK theory describes the transmission in a system of
$N$ parallel interacting disordered channels. On a formal level this brings us into the subject of
probability distribution over time--ordered products of random matrices of size $N\times N$
belonging to some noncommutative noncompact group $G$. This ``dynamic'' approach deals with the
construction of a Fokker--Planck equation for the eigenvalue distribution of random walks on the
group $G$ \cite{beenacker,pichard}. The DMPK Fokker--Planck equation can be regarded as a heat
equation for a radial part of Laplacian acting in the negative curvature symmetric coset space
$X^-=G/K$, where $K$ is the maximal compact subgroup of $G$ \cite{math,perelomov,caselle}.

The standard procedure implies the following steps. Introduce the radial coordinate $x_i$ (system
of roots) in the Cartan decomposition of $X^-$. Note that all $x_i$ ($i=1,...,N$) are ordered:
$x_1<x_2<...<x_N$. The distribution of $x_i$ $(i=1,...N)$ satisfies the heat equation
\be
\frac{\partial}{\partial t}W(\{{\bf x}\},t) = D \Delta_{\rm BL}^{(\rm r)} W(\{{\bf x}\},t),
\label{eq:48}
\ee
where $D$ is the diffusion coefficient and $\Delta_{\rm BL}^{(\rm r)}$ is the radial part of
Beltrami--Laplace operator in the symmetric space $X^-$. The explicit form of $\Delta_{\rm
BL}^{(\rm r)}$ acting in $X^-$ is well known -- see, for example, \cite{perelomov, perelomov2,
caselle2}. Define by $R=\{\alpha\}$ the root space of the Lie algebra $A_{N}$. Let $m_{\alpha}$ be
the multiplicity of the root $\alpha$ and consider the case $m_{\alpha}=\beta$ (for short roots)
and $m_{\alpha}=1$ (for long roots). We denote by $R^+$ the subspace of roots in $R$ that is
positive with respect to the Weyl chamber of $R$. This allows one to write $\Delta_{\rm BL}$ as
\be
\Delta_{\rm BL}^{(\rm r)}=\xi^{-2}\{{\bf x}\}\sum_{k=1}^N \frac{\partial}{\partial x_k} \xi^2\{{\bf
x}\} \frac{\partial}{\partial x_k},
\label{eq:49}
\ee
where
\be
\xi\{{\bf x}\} = \prod_{\alpha\in R_+} \left(\sinh x_{\alpha}\right)^{\nu_\alpha}= \prod_{i<j}
\left|\sinh^2 x_j-\sinh^2 x_i\right|^{\beta/2} \prod_i |\sinh 2 x_i|^{1/2}.
\label{eq:50}
\ee
One sees that the expression \eq{eq:49} for $N=1$ gives the radial part of Beltrami--Laplace
operator in the Lobachevsky plane $H_2$, where the single root, $\mu_1$ of the Cartan algebra is
the geodesic distance on a pseudosphere. Recall that for $N=1$ the Liouville--type equation
\eq{eq:21} coincides with the random walk in Lobachevsky plane.

The substitution $W(\mu_1,...,\mu_N,t)= \xi(\mu_1,...,\mu_N) \Psi(\mu_1,...,\mu_N,t)$ maps the
equations \eq{eq:48}, \eq{eq:49} onto the Schr\"odinger equation
\be
-\frac{\partial}{\partial t} \Psi({\mu},t) = H \Psi({\mu},t),
\label{eq:52}
\ee
with $H$ of hyperbolic Calogero--Sutherland--type form
\be
H=-\frac{1}{2}\sum_j \left(\frac{\partial^2}{\partial x_j^2} + \frac{1}{\sinh^2 2 x_j} \right) +
\frac{\beta(\beta-2)}{4} \sum_{j<k} \left(\frac{1}{\sinh^2(x_k-x_j)} +
\frac{1}{\sinh^2(x_k+x_j)}\right) + c,
\label{eq:53}
\ee
where $c$ is some irrelevant constant.

It has been pointed out in \cite{caselle2} that the identification of \eq{eq:49}--\eq{eq:53} with
the original DMPK equation \cite{beenacker} is possible if and only if the coupling constant in the
Calogero--Sutherland Hamiltonian, $g^2=\frac{\beta(\beta-2)}{4}$, equals to $\frac{1}{8}m_{\alpha}
(m_{\alpha}-2)|\alpha|^2$, where $|\alpha|$ is the root length. While we remind that
$m_{\alpha}=\beta$ (for short roots), $m_{\alpha}=1$ (for long roots), in our further consideration
we shall not fix specific value of $\beta$, but consider it as an arbitrary parameter. This enables
us to establish the connection between hyperbolic Calogero--Sutherland \eq{eq:52}--\eq{eq:53} and
Toda \eq{eq:c1} in the so-called Inozemtsev limit \cite{inozem}.

To take the Inozemtsev limit in \eq{eq:53}, we should rescale the constant $\beta$ simultaneously
changing of coordinates $x_j$ $(j=1,...,N)$:
\be
\left\{\begin{array}{l}
\disp \beta= -\sqrt{2\kappa}e^{\Delta} \medskip \\
x_j = -\mu_j + j \Delta
\end{array}\right. \qquad (\Delta \to \infty).
\label{eq:54}
\ee
Applying \eq{eq:54} to \eq{eq:53} and denoting the limiting Hamiltonian by $\tilde{H}$, we get
\be
-2\tilde{H}=\frac{1}{2}\sum_j \left(\frac{\partial^2}{\partial \mu_j^2} + \kappa e^{2(x_j-x_{j+1})}
\right),
\label{e:c2}
\ee
which coincides with the Hamiltonian of Toda chain.

Let us end up this Section by highlight the key observation. i) On one hand, the transmission in
multi--channel disordered wires, which corresponds to the Calogero--type system for a special
choice of coupling constant, can be described by diffusion in a symmetric space. ii) On the other
hand, the diffusion equation in the noncompact space $H_N=SL(N,R)/SO(N)$ corresponds to Toda--type
system and appears in the consideration of sequential uniform ballistic deposition in multi--column
box. The Inozemtsev limit \cite{inozem} allows to establish the connection between these two
different problems i) and ii).

\section{Two views on Lax representation for periodic Toda chain}
\label{sect:4.1}

In this Section we shall briefly consider the NNN ballistic deposition problem with periodic
boundary conditions. We have already argued before that the problem is described by the Toda chain
system. However we did not pay much attention on the boundary conditions. This is an important
issue since the periodic and open Toda chains behave quite differently. It is known that the open
chain is related to the motion at $SL(N,R)/SO(N)$ while the periodic chain is related to the motion
at the homogenous space of the affine algebra. Moreover the classical motion in the periodic case
is described in terms of the Riemann surfaces whose Jacobians are identified with the complex
Liouville tori of the dynamical system. The solution to the classical equation in the periodic case
is described via the elliptic functions at the Riemann surface while the quantum problem
responsible for the NNN ballistic deposition problem is quite involved.

Let us make a couple of general comments concerning the Toda realization of the BD model discussed
above with the periodic boundary conditions. First, consider the classical Toda model which allows
two different Lax representations. If there are $N$ degrees of freedom corresponding to the motion
at $SL(N,R)/SO(N)$, the Lax operator can be written either as $2\times 2$ or $N \times N$ matrix.
In the first representation the Lax operator in terms of the conventional phase space coordinates
reads as
\be
L_{\rm Toda}=\left( \begin{array}{cc} \lambda + p & e^q \\ -e^{-q} & 0
\end{array} \right).
\ee
This Lax operator acts as the shift operator on the Baker--Akhiezer (BA) function along the sites
of the chain. To compare with the discussion above consider the two--component Baker--Akhiezer
function $\Psi_n= \left(\begin{array}{c} \psi_n \\ \chi_n \end{array}\right)$. Then the linear
problem  can be rewritten as
\be
\psi_{n+1}-p_n\psi_n+e^{q_n-q_{n-1}}\psi_{n-1}=\lambda\psi_n,\quad \chi_n=-e^{q_{n-1}}\psi_{n-1}.
\ee
Putting $\chi_n = {\rm const}$ and freezing the dynamical variables in the system we get the
equation equivalent to \eq{eq:8}. Thus, we come to the following important conclusion:

\noindent{\bf Statement.} {\em The first component of the Baker--Akhiezer function $\Psi_n$ plays the
role of the finite--temperature partition function of a directed polymer in the random environment.
The local heights can be extracted from $\Psi_n$ as in \eq{eq:9} by taking logarithm and passing to
zero--temperature limit.}

This statement approves the link between the NNN ballistic deposition and the ``soft'' deposition
conjectured in the Section \ref{sect:2.3}.

Note that the $2\times 2$ Lax operator plays the role of the Hamiltonian for the discrete fermionic
system while \eq{eq:5} can be identified with the transfer--matrix of Toda model.

The solution to the classical equations of motion is described by the spectral curve
\be
\det\big[T(\lambda)- \omega\big]=0; \qquad T(\lambda)=\prod_{i=1}^{N} L_i(\lambda),
\ee
where variable $\omega$ corresponds to the quasimomentum for periodic boundary conditions.
Generically it is genus $(N-1)$ curve which gets degenerated into the rational one for the
non-periodic case. This spectral curve is the Liouville torus for the dynamical system whose
degrees of freedom are localized on this curve. In terms of the spectral curve one can define the
separated variables that constitute the set of identical noninteracting degrees of freedom. These
degrees of freedom are identified with poles of the BA function on the spectral curve.

The question deserving special attention concerns the possibility to find a Douglas--Kazakov phase
transition \cite{douglas} in a Toda chain for different boundary condition since this transition is
responsible for Kardar--Parisi--Zhang scaling (see \cite{maj} for details). We plan to discuss this
question in a separate publication and as a preliminary remark we outline in Appendix \ref{app:2}
the connection between solutions of 1D Toda lattice equation and rational solutions of Painlev\'e
II.

\begin{figure}[ht]
\epsfig{file=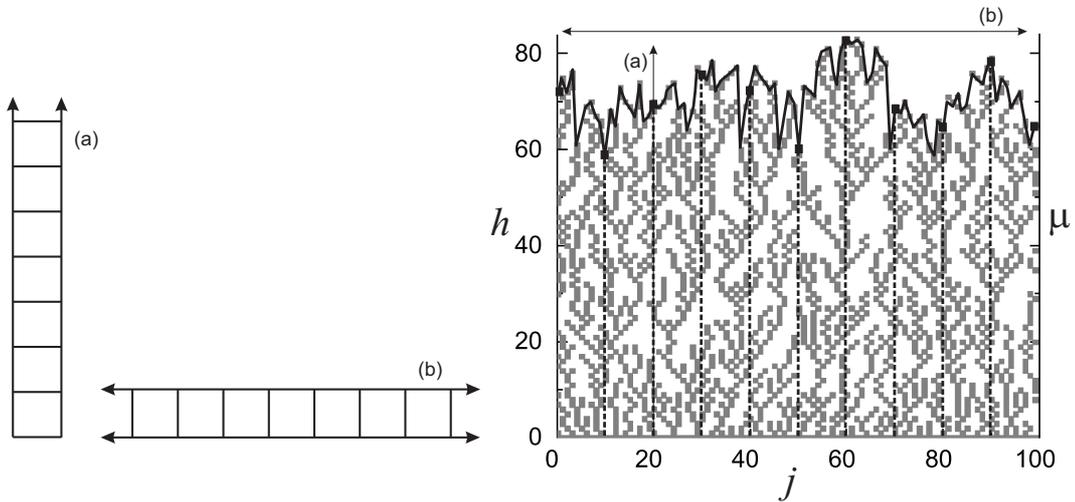,width=14cm} \caption{Two possibilities of constructing Lax representation
for Toda chain: (a) ``time--like'' and (b) ``space--like''. The visualization of these
representations in terms of growing heap is shown nearby.}
\label{fig:space-time}
\end{figure}

The Toda system can be also written starting from the linear problem
\be
{\cal L}= \left(\begin{array}{cccccc} -p_1 & e^{{1\over 2}(q_2-q_1)} & 0 & 0 & \hdots & {1\over
w}e^{{1\over 2}(q_{n}-q_1)} \medskip \\
e^{{1\over 2}(q_2-q_1)} & -p_2 & e^{{1\over 2}(q_3-q_2)} & 0 & \hdots & 0 \medskip \\
0 & e^{{1\over 2}(q_3-q_2)} & -p_3 & e^{{1\over 2}(q_4-q_3)} & \hdots & 0 \medskip \\
0 & 0 & e^{{1\over 2}(q_4-q_3)} & -p_4 & \hdots & 0 \medskip \\
\vdots & \vdots & \vdots & \vdots & \ddots & \vdots \medskip \\
-{w}e^{{1\over 2}(q_{n}-q_1)} & 0 & 0 & 0 & \cdots & -p_n
\end{array} \right),
\label{eq:L_toda}
\ee
with the $N$--component Baker--Akhiezer function $\Phi=\{e^{-q_n/2}\psi_n\}$. This leads us to the
spectral curve
\be
\det_{n\times n}\left({\cal L}(w)-\lambda\right)=0,
\label{hren}
\ee
which is equivalent to the spectral curve in the $2\times 2$ representation and reads as
\be
\omega +\omega^{-1}=P_{N}(\lambda).
\label{eq:omega}
\ee
The symmetry of two different representations of Lax operator by $2 \times 2$ or $N \times N$
matrices can be interpreted as ``time''--like (a) and ``space''--like (b) representations
schematically depicted in \fig{fig:space-time}. Geometrically this symmetry means the rotation of
the picture in the homogeneous space -- see \cite{ggm} for details. In terms of ballistic
deposition the time--like representation can be visualized as growth along the time running since
the first deposition event, while the space--like representation can be associated with growth
``along a surface'' of the heap (i.e. in the direction orthogonal to the time axis).

Let us conclude the reminder about two types of Lax representations by noting that ideologically
similar description is allowed for the one--dimensional Anderson model. Two different points of
view on this problem, described below, deal nothing with integrability, however they have exactly
the same physical interpretation as two different Lax representations of 1D Toda chain. Following
\cite{haake}, consider the electron hopping on the one--dimensional lattice described by a
tight--binding one--dimensional Schr\"odinger equation with site disorder. Denote by $\psi_j$ and
$U_j$ the wave function and the random potential on the lattice site $j$. Then the eigenvalue
problem is given by the equation
\be
\psi_{j-1}+ U_j\psi_j(x)+\psi_{j+1}=E\psi_j(x).
\label{eq:45}
\ee
Equation \eq{eq:45} admits two representations. Introducing the two--component vector
$(\psi_{j+1},\psi_j)$, we can rewrite \eq{eq:45} as
\be
\left(\begin{array}{c} \psi_{j+1} \\ \psi_j \end{array}\right) = L_j \left(\begin{array}{c} \psi_j \\
\psi_{j-1} \end{array}\right); \qquad L_j=\left(\begin{array}{cc} E-U_j & -1
\\ 1 & 0 \end{array}\right),
\label{eq:46}
\ee
which has similarity with \eq{eq:L_toda}. From the other hand, Eq. \eq{eq:45} has representation in
terms of spectral problem for $N\times N$ tridiagonal matrix ${\cal L}$ applied to the vector
$\Psi=\{\psi_1,\psi_2,...\psi_N\}$:
\be
{\cal L}=
\left(\begin{array}{cccccc} E-U_1 & -1 & 0 & 0 & \hdots & a_{1n} \medskip \\
-1 & E-U_2 & -1 & 0 & \hdots & 0 \medskip \\
0 & -1 & E-U_3 & -1 & \hdots & 0 \medskip \\
0 & 0 & -1 & E-U_4 & \hdots & 0 \medskip \\
\vdots & \vdots & \vdots & \vdots & \ddots & \vdots \medskip \\
a_{n1} & 0 & 0 & 0 & \cdots & E-U_N
\end{array} \right).
\label{eq:47}
\ee
It might be interesting to check whether such a structural analogy between two Lax representations
of Toda chain and two descriptions of Anderson model has more deep physical grounds than just a
formal similarity.

\section{Conclusion}
\label{sect:5}

In this paper we have reformulated the NNN ballistic deposition process in terms of time--ordered
product of the generators of a local group $L_N$ (see the Definition 1). The generators of this
group are the $N\times N$--matrices consisting of a single $sl_2$--block randomly located along the
diagonal. We have related the Markov dynamics on $L_N$ with the diffusion in the metric space
$H_N=SL(N,R)/SO(N)$.

Let us summarize the main results of our work:

\begin{itemize}

\item By constructing the Beltrami--Laplace operator on $H_N$ in an appropriate system of
coordinates obtained via Gauss decomposition, we claim the existence of the integrable structure
behind the NNN deposition problem.

\item We demonstrate the relation between the NNN growth and dynamics of the 1D Toda chain where
the number of sites in the chain coincides with the number of the sites in the deposition problem.
The largest geodesic distance in the corresponding symmetric space $H_N=SL(N,R)/SO(N)$ is
identified with the maximal local heights in the deposition process.

\item Taking the Inozemtsev limit in the hyperbolic Calogero--Sutherland model we uncover the
connection between the ballistic deposition process and the transmission in multi--channel random
wires. This connection certainly deserves further study.

\end{itemize}

It is worth pointing out the relation between the fluctuations of the partition function of a
directed polymer at finite temperatures and the quantum Toda lattice has been obtained by N.
O'Connell in a model of a semi--continuous directed polymer model with the length related to
increments of Brownian motions \cite{oconnel}. Within this model it was obtained that the free
energy, as a function of $t$, behaves like the first component of a process in $\mathbb R^N$ which
also satisfies a diffusion equation with a Toda Hamiltonian. The extension of this work to
statistics of vicious random walks with "a long--ranged killing term" has been considered in a
recent work \cite{katori}. It would be very interesting to establish correspondence between the
subject of our work, the problems analyzed in \cite{oconnel, katori} and eventually to the
Yang--Mills formulation of the statistics of vicious random walks developed recently in \cite{maj}.

Let us outline the possible directions for further investigation. It seems to us that the most
interesting case corresponds to the periodic boundary conditions. In this case the deposition
problem gets mapped onto the periodic Toda chain whose solution to the classical equation of motion
is described by the Riemann surface. In this case one could define the prepotential which yields
the solution of the dispersionless Whitham hierarchy similar to the one considered in \cite{gmmm}.
We plan to discuss the periodic case separately.

Note the analogy between the ballistic deposition and the Laplacian growth (LG). In LG case the
integrable system is defined as well, where the natural degrees of freedom are attributed to the
shape of the droplet while the Whitham times get identified with the momenta of the droplet. The
initial evolution equation selects the particular solution of the Whitham hierarchy. In the BD case
the local heights play the role of the Toda degrees of freedom. From the physical point of view the
difference between the LG and BD consists only in deposition rules: in Laplacian growth the
particles follow Brownian trajectories until they are attached to the cluster, while in ballistic
depositions the particles follow straight parallel trajectories as in a rain. The passage from LG
to BD consists in ``elongation'' of Brownian trajectories in an external field. This point of view
is the subject of our forthcoming work.

Our final remark concerns the existence of shocks in Toda chain \cite{kamv}. In \cite{shocks} we
have demonstrated that the discrete stochastic equation describing the BD process in the NNN model
can be naturally represented in terms of a dynamic programming language. This dynamic programming
point of view allowed for a systematic translation of the study of growing clusters and channels
separating them (``crevisses'')  into the language of maximizers and shocks in discrete equations
of Burgers or Hamilton--Jacobi type. Whether these geometrical properties of a growing heap have
interpretation in terms of shocks in Toda chain, is an open and challenging question.

\bigskip

\centerline{\bf Acknowledgments}

The authors are grateful to A. Comtet, Y. Fyodorov, S. Majumdar and V. Priezzhev for encouraging
discussions. The work of A.G. was partially supported by grants RFBR-09-02-00308 and
CRDF-RUP2-2961-MO-09. A.G. is grateful to Simons Center for Geometry and Physics and KITP at UCSB,
where part of this work has been done, for the hospitality and support.

\begin{appendix}

\section{The connection between the columnar heights and coordinates in the symmetric space}
\label{app:1}

1. We expect for $N\ge 2$ the following linear relation between heights $\{h_1,...,h_N\}$ and
geodesic distances $\{\mu_1,...,\mu_N\}$ to hold:
\be
\max \{h_1,...,h_N\} = \gamma_{N} \max \{\mu_1,...,\mu_N\}
\label{eq:max}
\ee
where $\gamma_N$ is some constant, $\mu_j=\ln \lambda_j$ ($j=1,...,N$) and $\lambda_j$ are the
eigenvalues of the matrix $\tilde{V}_t$, which is the product of generators of locally free
semigroup $F^+_N$. The behavior \eq{eq:max} is supported by numerical simulations of NNN uniform
ballistic deposition in a $N$--columnar bounding box, considered as: i) sequential adding of $T$
elementary blocks with NNN interactions, and ii) time--ordered multiplication of $T$
matrices--generators of the locally free semigroup $F_N^+$.

\begin{figure}[ht]
\epsfig{file=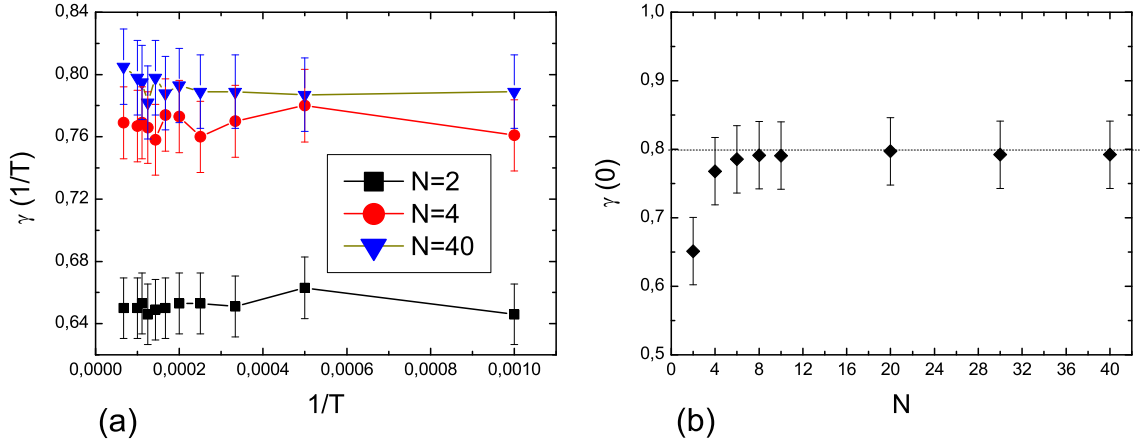,width=15cm}
\caption{Linear dependence of maximal height for NNN--deposition on the logarithm of the
maximal eigenvalue in corresponding matrix representation: a) dependence of $\gamma_N$ on $1/T$ for
fixed $N$; b) dependence of $\gamma_N(0)$ on different $N$.}
\label{fig:hdist}
\end{figure}

In the \fig{fig:hdist} (a) we plot the ratio $\gamma_N(T)=\max \{h_1,...,h_N\}/\max
\{\mu_1,...,\mu_N\}$ in the interval $T=1000 \div 10\ 000$ for fixed values of $N$ (for better
visibility the coordinates $(\gamma_N, 1/T)$ are used). We see that for each $N$ the coefficient
$\gamma_N(1/T)$ tends for $T\to \infty$ to some selfaveraged value $\gamma_N(0)$.

The figure \fig{fig:hdist} (b) demonstrates the dependence of $\gamma_N(0)$ on $N$ for $N=$ 2, 4,
6, 8, 10, 20, 30, 40. The value $\gamma_N(0)$ saturates very rapidly with increasing of $N$ and
reaches the limiting value $\approx 0.8$.

The results of simulation depicted in \fig{fig:hdist} demonstrate the linear dependence of the
maximal height on the logarithm of the largest eigenvalue in associated matrix representation.
Moreover, for $N\gg 1$ the coefficient $\gamma_N$ becomes independent on $N$.

\bigskip

2. The relation between the height in $N=2$--ballistic deposition and the corresponding hyperbolic
distance in $H_2=SL(2,R)/SO(2)$ can be easily established. To see that, note that the height in the
column is in one--to--one correspondence with the shortest distance on the corresponding Cayley
graph -- see \fig{fig:two_three} of the Section \ref{sect:2.3}. Let us discuss in details the
two--columnar deposition and show that the maximal height in two neighboring columns, i.e. the
average number of steps along a Cayley graph is linearly proportional to the hyperbolic distance,
$d$, in the space $H_2$. We use some facts about the isometric embedding of graphs in hyperbolic
planes and uniform Markov processes in hyperbolic domains -- for more details see, for example,
\cite{cnv}.

\begin{figure}[ht]
\epsfig{file=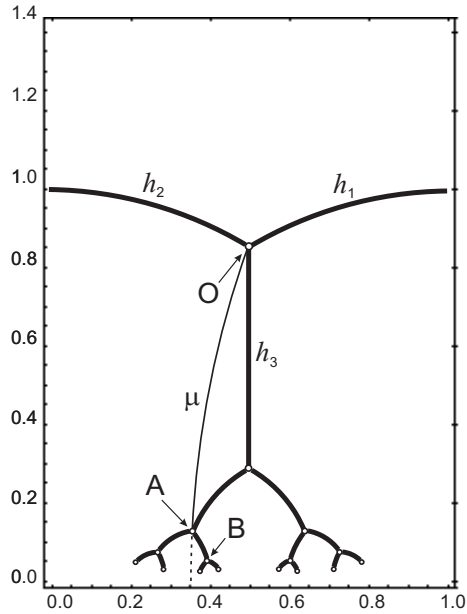, width=6cm}
\caption{Poincar\'e hyperbolic upper half--plane $H_2$. $A$ and $B$ are two neighboring
vertices of Cayley tree isometrically embedded into $H_2$; $d$ is the hyperbolic distance between
points $A$ and $B$.}
\label{fig:appendix1}
\end{figure}

Consider the hyperbolic Poincar\'e upper half--plane $H_2=\{z \in C,\,{\rm Im}(z)>0\}$. The matrix
representation of the generators of the group $\Gamma_2$ is well known (see, for example
\cite{terras}), however for our purposes it is more  convenient to take a basis consisting of the
composition of standard fractional--linear transform and complex conjugacy. Namely, denoting by
$\bar{z}$ the complex conjugate of $z$, we consider the following action
\be
\left(\begin{array}{cc} a & b \\ c & d \end{array} \right):\ z\to\frac{a\bar z+b}{c\bar z+d}
\ee
A possible set of generators is then:
\be \label{hk}
h_0=\left(\begin{array}{cc} 1 & -2/\sqrt{3} \\ 0 & -1 \end{array} \right),\;
h_1=\left(\begin{array}{cc} 1 & 2/\sqrt{3} \\ 0 & -1 \end{array} \right),\;
h_2=\left(\begin{array}{cc} 0 & 1 /\sqrt{3} \\ \sqrt{3} & 0 \end{array} \right)
\end{equation}
Choosing the point $(x_0,iy_0)=(0,i)$ as the tree root---see fig.\ref{fig:appendix1}, any vertex on
the tree is associated with an element $\disp V_n=\prod_{k=1}^n h_{\alpha_k}$ where
$\alpha_k\in\{0,1,2\}$ and is parameterized by its complex coordinates $z_n=V_{n}\big((-1)^n
i\big)$ in the hyperbolic plane.

For $H_2=SL(2,R)/SO(2)$ one can identify an element with its class of equivalence of $SO(2)$. If
one denotes by $\mu(V_n)\equiv \mu(i,z_n)$ the hyperbolic distance between $i$ and $z_n$, the
following identity holds
\be
2\cosh \mu(V_n)={\rm Tr}(V_{n}V_{n}^{\dag})
\label{eq:cosh}
\ee
where dagger denotes transposition.

Let $W_n(\mu)$ be the probability to find the tree vertex in the generation $n$ at the hyperbolic
distance $\mu$ from the root point. In terms of deposition process, $n$ is the average height of
two neighboring columns). It means that we are looking for the distribution of the traces for
matrices $V_n$ which are irreducible products of $n$ sequential multiplications. If we denote by
$\mu(V_n)$ the irreducible length of the word represented by the matrix $V_n$, then $V_n$ is
irreducible if and only if $\mu(V_n)=n$. Such a word enumeration problem is simple in case of the
group $\Gamma_2$, because of its free product structure: $\Gamma\sim Z_2\otimes Z_2\otimes Z_2$.
Indeed, if $\disp V_n= \prod_{k=1}^n h_{\alpha_k}$ one has $\mu(V_n)=n$ if and only if
$h_{\alpha_k}\not= h_{\alpha_{k-1}}\; \forall k$. Hence we have to study the behavior of the random
matrix $V_n$, generated by the following Markovian process
\be
V_{n+1}=V_n h_{\alpha_{n+1}}\ {\rm with}\ \alpha_{n+1}=\left\{\begin{array}{ll}
(\alpha_{n}+1)\,{\rm mod}\,3 & \ \mbox{with probability $\frac{1}{2}$}
\medskip \\ (\alpha_{n}+2)\,{\rm mod}\,3 & \  \mbox{with probability $\frac{1}{2}$}
\end{array} \right.
\label{mar}
\ee

One can use the standard methods of random matrices and consider the entries of the $2\times
2$--matrix $V_n$ as a 4--vector ${\cal V}_n$. The transformation $V_{n+1}= V_n h_{\alpha}$ reads
\be
{\cal V}_{n+1}= \left(\begin{array}{ll} h_{\alpha}^{\dag} & 0 \\ 0 & h_{\alpha}^{\dag}
\end{array}\right)\; {\cal V}_{n}
\ee
This block--diagonal form allows one to restrict ourselves to the study of one of two 2--vectors,
composing ${\cal V}_{n}$, say $v_n$. Parameterizing $v_n=(\varrho_n\cos\theta_n, \varrho_n\sin
\theta_n)$ and using the relation $\mu(V_n)\equiv \mu_n\simeq 2\ln\varrho_n$ valid for $n\gg 1$,
one gets a recursion relation $v_{n+1}=h_{\alpha}^{\dag}v_n$ in terms of hyperbolic distance
$\mu_n$:
\be
\disp  \mu_{n+1}= \mu_n+ \ln\, \left[\frac{5}{3}+\frac{4}{3}\cos(2\theta_n+\varphi_{\alpha})\right]
\label{dist}
\ee
where $\varphi_{\alpha}$ depends on the transform $h_{\alpha}$ through $\disp \varphi_{\alpha}
=(2\alpha-1)\pi/3\; (\alpha=0,1,2)$, while for the angles one gets straightforwardly
\be
\tan\theta_{n+1}=h_{\tilde{\alpha}}(\tan\theta_n)
\label{ang}
\ee
and the change $\alpha\to \tilde{\alpha}$ means the substitution $(0,1,2)\to (1,0,2)$. Action of
$h_{\alpha}$ is still fractional--linear. The properties of the invariant measure
$\nu_{\alpha}(\theta)$ have been discussed by Gutzwiller and Mandelbrot \cite{gut}. We are now led
to compute the Lyapunov exponent, $\gamma$ which is
\be
\gamma=\lim_{n\to\infty} \frac{\left<\mu\right>}{n}=
\int_{0}^{\frac{\pi}{3}}d\theta\mu_1(\theta-\frac{\pi}{6})
\ln\left(\frac{5}{3}+\frac{4}{3}\cos2\theta\right)
\label{gamma}
\ee
The numerical value of $\gamma$ can be obtained by means of semi--numerical procedure which
involves the numerical information about the invariant measure $\mu_1(\theta)$ \cite{cnv,gut}. One
finally gets: $\gamma\approx 0.79$. By this we prove the linear relation $\mu=\gamma n$ between
hyperbolic distance, $\mu$, in $H_2$ and average height in two neighboring columns, $n$.

\section{Minimal action principle for Toda system and rational solutions of Painlev\'e II}
\label{app:2}

Represent the partition function $W$ in the form of Feynmann--Kac path integral with the Lagrangian
obtained by Legendre transformation from Toda Hamiltonian \eq{eq:38}:
\be
W(\mu_1,...,\mu_N,t)=\int ... \int {\cal D}\{\mu\}
\exp\left\{-\int\limits_{0}^{t}ds\left[\sum_{j=1}^N\dot{\mu}_j^2(s)+V\Big(\mu_j(s),\mu_{j+1}(s)\Big)
\right]\right\},
\label{eq:55}
\ee
where $V(\mu_j,\mu_{j+1})= e^{\mu_j-\mu_{j+1}}$ is the Toda potential. Minimizing the corresponding
action, we arrive at the standard Euler equations, which describe the dynamics of 1D Toda chain:
\be
\ddot{\mu}_j(s) = e^{\mu_j-\mu_{j+1}}-e^{\mu_{j+1}-\mu_{j+2}},
\label{eq:56}
\ee
where $\ddot{\mu}_j(s)\equiv \frac{d^2}{d s^2} \mu_j(s)$. There are many equivalent representations
of Toda chain equation \eq{eq:56}. Among the most known, is the bilinear form of \eq{eq:56} (see,
for example, \cite{toda-bilinear}):
\be
\tau_j'' \tau_j - (\tau_j')^2=\tau_{j+1}\tau_{j-1},
\label{eq:57}
\ee
where $\tau_j'= \frac{d}{ds} \tau_j$, $\tau_j''= \frac{d^2}{ds^2} \tau_j$ and
\be
\mu_j = \ln \frac{\tau_{j-1}}{\tau_j}.
\label{eq:58}
\ee
The tau--function $\tau_n$ can be written in the determinant form (see, for example,
\cite{toda-det})
\be
\tau_j=\det\left(\begin{array}{cccc}
\varphi & \frac{d}{ds} \varphi & \cdots & \left(\frac{d}{ds}\right)^{j-1} \varphi \medskip \\
\frac{d}{ds} \varphi & \left(\frac{d}{ds}\right)^{2} \varphi & \cdots &
\left(\frac{d}{ds}\right)^{j} \varphi \medskip \\ \vdots & \vdots & \ddots & \vdots \medskip \\
\left(\frac{d}{ds}\right)^{j-1} \varphi & \left(\frac{d}{ds}\right)^{j} \varphi & \cdots &
\left(\frac{d}{ds}\right)^{2j-2} \varphi \\
\end{array} \right).
\label{eq:59}
\ee
The function $\varphi$ for open boundary conditions is an arbitrary function of $s$. Applying a
gauge transformation to $\tau_j$
\be
\frac{d^2}{ds^2} \ln \tau_j(s) = \frac{d^2}{ds^2} \ln \sigma_j(s) + p q,
\label{eq:60}
\ee
with arbitrary $p$ and $q$, one arrives at the equation
\be
\sigma_j'' \sigma_j - (\sigma_j')^2=\sigma_{j+1}\sigma_{j-1}-pq\sigma_j^2,
\label{eq:61}
\ee
where $\sigma_{-1}=p$, $\sigma_0=1$, $\sigma_1=q$. Equation \eq{eq:61} resembles the
differential--difference equation that defines Yablonskii--Vorob'ev polynomials \cite{yablonskii1},
$Q_n(z)$:
\be
Q_j''(z)Q_j(z) - (Q_j'(z))^2= - \frac{1}{4}Q_{j+1}Q_{j-1} + \frac{z}{4} Q_j^2(z),
\label{eq:62}
\ee
with $Q_0=1$, $Q_1=z$ and the derivatives taken with respect to $z$. To establish the precise
connection between \eq{eq:61} and \eq{eq:62}, let us make the shift in \eq{eq:61}: $\sigma_j \to
A^{j^2} \sigma_j$ and then fix the values of $A$, $p$, and $q$ to satisfy conditions:
$A=\frac{i}{2}$, $pq=-\frac{z}{4}$.

It is known \cite{rational,yablonskii2} that the Yablonskii--Vorob'ev polynomials $Q_n(z)$ are
connected to the rational solutions of Painlev\'e II. Namely, the combination
\be
w_j(z) = \frac{d}{dz} \ln \frac{Q_{j-1}(z)}{Q_j(z)} \equiv \frac{Q_{j-1}'(z)}{Q_{j-1}(z)}-
\frac{Q_j'(z)}{Q_j(z)}
\label{eq:63}
\ee
gives the rational solutions of the Painlev\'e II equation
\be
w''(z)=2w^3(z)+ z w(z)+ \alpha \qquad (\alpha = j \in \mathbb{Z}).
\label{eq:64}
\ee
So, the classical optimal trajectory solution for Toda Hamiltonian can be linked by equations
\eq{eq:55}--\eq{eq:64} to the rational solutions of the Painlev\'e II equation.

\end{appendix}

\end{document}